\documentclass[twocolumn,prd,aps,showpacs,showkeys,amsmath,amssymb]{revtex4-1}
\usepackage{bm}

\usepackage{color}
\usepackage{amsmath}
\usepackage{amsfonts}
\usepackage{verbatim}
\usepackage{amssymb}
\usepackage{graphicx}
\usepackage{epstopdf}
\usepackage{mathrsfs}
\usepackage{epsfig}
\usepackage{slashed}
\usepackage{bbold}

\providecommand{\U}[1]{\protect\rule{.1in}{.1in}}
\newcommand{\bl}{\boldsymbol}

\newcommand{\ph}{\phantom}

\newcommand{\eq}{\,=\,}
\newcommand{\ma}{\,+\,}

\begin{document}

\title{A Class of Integrable Metrics and Gauge Fields}
\author{Gabriel Luz Almeida and Carlos Batista}
\email[]{carlosbatistas@df.ufpe.br}
\affiliation{Departamento de F\'{\i}sica, Universidade Federal de Pernambuco,
Recife, Pernambuco  50740-560, Brazil}


\begin{abstract}
Starting with the most general four-dimensional spacetime possessing two commuting Killing vectors and a nontrivial Killing tensor, we analytically integrate Einstein-Yang-Mills equations for a completely arbitrary gauge group. It is assumed that the gauge field inherits the symmetries of the background and is aligned with the principal null directions of the spacetime.
\end{abstract}
\keywords{Einstein-Yang-Mills, Exact Solutions, Gauge Fields}
\maketitle

\section{Introduction}

Almost four decades ago, S. Benenti and M. Francaviglia have been able to find the most general $N$-dimensional spacetime endowed with $(N-2)$ commuting Killing vector fields and a nontrivial Killing tensor \cite{BenentiFrancaviglia}. Particularly, in four dimensions, $N=4$, the metric of this spacetime, written in the inverse form, is given by:
\begin{align}
 g^{ab}\partial_{a}\partial_{b}  =&\frac{1}{S_{1}(x)+S_{2}(y)}\, \Big[
\,G_{1}^{ij}(x)\,\partial_{i}\partial_{j}\,-\,G_{2}^{ij}(y)\,\partial_{i}\partial_{j}    \nonumber\\
& \quad \quad \quad +   \Delta_{1}(x)\,   \partial_{x}^{2}+\Delta_{2}(y)\, \partial_{y}^{2}\, \Big]  \, , \label{BFmetric}
\end{align}
where in the above expression the indices $a,b$ range  over the coordinates $\{\tau,\sigma,x,y\}$, whereas the indices $i,j$ run just over the cyclic coordinates $\{\tau,\sigma\}$. The functions $G_{1}^{ij}$, $S_1$ and $\Delta_{1}$ are arbitrary functions of $x$ whereas $G_{2}^{ij}$, $S_2$ and $\Delta_{2}$ are arbitrary functions of $y$, with $G_{1}^{ij}$ and $G_{2}^{ij}$  being symmetric on the indices $i,j$, comprising a total of 10 free functions on the metric. However, note that by redefining the coordinates $x$ and $y$ we can easily get rid of two of these functions. For instance, we could set $\Delta_{1}(x)=1$ and $\Delta_{2}(y)=1$, although we will not make such a choice at this point. Besides the obvious Killing vectors $\partial_\tau$ and $\partial_\sigma$, the spacetime with metric \eqref{BFmetric} is also endowed with the following rank two Killing tensor
\begin{align}
  \boldsymbol{K}\,=\,\frac{-1}{S_{1}+S_{2}}\,\Big[& \,S_{1}\,G_{2}^{ij}\, \partial_{i}\partial_{j}
+ S_{2}\,G_{1}^{ij}\,\partial_{i}\partial_{j}   \nonumber\\
& + \Delta_{1} \,S_{2} \,  \partial_{x}^{2}-S_{1}\,\Delta_{2} \,\partial_{y}^{2} \, \Big]  \,. \label{KillingT1}
\end{align}
Since the metric is covariantly constant, it is also a rank two Killing tensor. Therefore, the geodesic motion of the spacetime with metric \eqref{BFmetric} has four first integrals, two that are linear on the momentum, constructed from the Killing vectors, and two that are quadratic on the momentum stemming from the Killing tensors. Thus, the geodesic motion is completely integrable for this class of spacetimes.

In the recent work \cite{AnabalonBatista}, A. Anabal\'{o}n along with one of us (C. Batista) attained the full integration of Einstein's vacuum equation for the general class of metrics of the form \eqref{BFmetric} under the following assumptions:
\begin{equation}\label{detF}
  \left.
     \begin{array}{ll}
       \det G_{1}^{ij}  \,\equiv\, G_{1}^{\tau\tau} G_{1}^{\sigma\sigma} - G_{1}^{\tau\sigma} G_{1}^{\tau\sigma} = 0 \,,\\
       \det G_{2}^{ij}   \,\equiv\, G_{2}^{\tau\tau} G_{2}^{\sigma\sigma} - G_{2}^{\tau\sigma} G_{2}^{\tau\sigma} = 0\,.
     \end{array}
   \right.
\end{equation}
The above constraints guarantee that the part $G_{1}^{ij} \partial_{i}\partial_{j}$ of the inverse metric tensor can be written as the square of a vector field, the same being true for the part $G_{2}^{ij} \partial_{i}\partial_{j}$, so that the spacetime has a naturally defined Lorentz frame and, consequently, a natural null tetrad. Particularly, this class of metrics contains Carter's line element as a special case \cite{DeWitt:1973uma}, which embraces some of the most physically relevant analytical solutions of Einstein's equation, such as Kerr-(A)dS spacetime.

The goal of the present work is to analytically integrate Einstein-Yang-Mills equations, for a completely arbitrary gauge group, for the class of spacetimes (\ref{BFmetric}), under the assumptions (\ref{detF}), extending the vacuum results of Ref. \cite{AnabalonBatista} to the case where the gravitational field interacts with a gauge field. During the integration process we will further assume that the gauge field is aligned with the principal null directions of the spacetime, a feature common to almost all known charged black hole solutions.

The nonlinearity of Einstein's equation precludes the attainment of a general analytical solution without the imposition of symmetries, which justifies our choice of working with the class of spacetimes that are stationary, axisymmetric, and with a nontrivial Killing tensor. In spite of the imposition of such symmetries, it is worth noting that our choice is not over-restrictive. Indeed, isolated astrophysical objects that attained the equilibrium state should be stationary and axisymmetric, as their energies and angular momenta should not change, a reasoning that is supported by the so-called rigidity theorem \cite{HawkingRigidity,Chrusciel:1996bj}. On the other hand, the Killing tensor, which represents a symmetry of the geodesic phase space \cite{Carter-KleinG,Santillan}, is required due to the desire of reaching full analytical integration.
In fact, since in the backgrounds considered here the dynamics of the geodesic motion is integrable, it seems natural to seek for the integrability of field equations in these backgrounds. As a matter of fact, Killing and Killing-Yano tensors, are behind several recent progresses in high energy physics, as exemplified by the integration of Klein-Gordon equation \cite{Frol-KG}, Dirac equation \cite{Oota-Dirac,KYoperator,CarterAngMom}, and gravitational perturbations \cite{OotaPerturb} in Kerr-NUT-(A)dS spacetimes in arbitrary dimensions. In addition, these hidden symmetries have shown to be of relevance in supersymmetric theories \cite{Gibbons_sky,KY-SUSY,Cariglia,KYString}.


Yang-Mills gauge theory is the main element behind the Standard Model, which is the most successful physical theory constructed so far, as it accurately models three of the four fundamental interactions in our Universe. Since the only missing interaction in this scheme is gravity, it is of great relevance to study how Yang-Mills fields interact with it. However, due to the intricate non-linearity of the non-abelian Einstein-Yang-Mills field equations, most solutions obtained so far are numerical \cite{Kleihaus:2000kg,Kuenzle:1990is,Bizon:1990sr,Kleihaus:1997ic,Radu:2002hf}, although some important analytical examples have been obtained \cite{Yasskin:1975ag,Hosotani:1984wj}, and even exact solutions for non-minimally coupled Yang-Mills fields have recently been attained \cite{Balakin:2016mnn}. Interestingly, in non-abelian theory a gauge field that is proportional to a pure gauge field does not have vanishing field strength, yielding the so-called meron solutions, which have no abelian analogue. Some analytical examples of solutions of the latter kind have also been found in the literature \cite{MeronsOliva}. Another important motivation for studying non-abelian gauge fields interacting with gravity is that the no-hair theorem is not valid in such a case \cite{Bartnik:1988am,Bizon:1990sr,Kleihaus:1997ic}, living room for a much richer arena of solutions compared with the abelian case \cite{Volkov_Review}.


The outline of the article is the following. In Sec. \ref{Sec.NullFrame} we show that the condition (\ref{detF}) leads to the existence of a natural null tetrad frame and show that this frame is endowed with relevant algebraic and differential properties. Then, in Sec. \ref{Sec.Metric}, the free functions of the metric (\ref{BFmetric}) are redefined in a way that is more suitable for the integration process. Next, in Sec. \ref{Sec.AlignementCondition} we investigate how the alignment condition constrains the components of the gauge field and show how the solution for the non-abelian problem can be obtained from the general solution for the case of an abelian gauge group, which comprises an important step for the attainment of our solutions. Then, in Sec. \ref{Sec.Abelian Case}, the integration of abelian Einstein-Yang-Mills equations is performed. Next, in Sec. \ref{Sec.General Gauge} the solution of the abelian problem is used to find the general solution for the non-abelian problem. In particular, the examples of the groups $SU(2)$ and $SO(4)$ are fully worked out. An overview and discussion of the results is done in Sec. \ref{Sec.Conclusions}. Finally, in Appendix \ref{AppendixCoord} we show the coordinate transformations that connect part of our solutions to some well-known solutions.


\section{The Natural Null Frame}\label{Sec.NullFrame}

If a $2\times 2$ matrix has vanishing determinant then its two columns are proportional to each other so that the matrix can be written as a tensor product of a column vector with itself. Therefore, assuming the constraints \eqref{detF}, it follows that we can write:
\begin{equation*}
  \left.
     \begin{array}{ll}
        G_{1}^{ij} \,\partial_{i}\partial_{j} = \left[ \, \sqrt{G_{1}^{\tau\tau} }\, \partial_\tau +  \sqrt{ G_{1}^{\sigma\sigma} } \,\partial_\sigma  \,\right]^2  \,,\\
       G_{2}^{ij} \,\partial_{i}\partial_{j} = \left[ \, \sqrt{G_{2}^{\tau\tau} }\, \partial_\tau +  \sqrt{ G_{2}^{\sigma\sigma} } \, \partial_\sigma  \,\right]^2 \,.
     \end{array}
   \right.
\end{equation*}
In such a case, the metric \eqref{BFmetric} is naturally written as the following sum of squares:
\begin{equation*}
  g^{ab}\partial_{a}\partial_{b}  = -(\bl{e}_0)^2 + (\bl{e}_1)^2 + (\bl{e}_2)^2 + (\bl{e}_3)^2 \,,
\end{equation*}
where the Lorentz frame $\{\bl{e}_0,\bl{e}_1,\bl{e}_2,\bl{e}_3\}$ is defined by
\begin{equation*}
  \left\{
     \begin{array}{ll}
       \bl{e}_0 = \frac{1}{\sqrt{S_1 + S_2}}\left( \sqrt{G_{2}^{\tau\tau} }\, \partial_\tau +  \sqrt{ G_{2}^{\sigma\sigma} } \, \partial_\sigma \right) \,, \\
       \bl{e}_1 = \frac{1}{\sqrt{S_1 + S_2}}\left( \sqrt{G_{1}^{\tau\tau} }\, \partial_\tau +  \sqrt{ G_{1}^{\sigma\sigma} } \, \partial_\sigma \right) \,, \\
       \bl{e}_2 = \frac{1}{\sqrt{S_1 + S_2}} \sqrt{\Delta_1}\,\partial_x  \,, \\
      \bl{e}_3 = \frac{1}{\sqrt{S_1 + S_2}} \sqrt{\Delta_2}\,\partial_y \,.
     \end{array}
   \right.
\end{equation*}
Out of this basis, we can immediately build a null tetrad frame $\{\bl{\ell},\bl{n},\bl{m},\overline{\bl{m}}\}$:
\begin{equation*}
  \left\{
     \begin{array}{ll}
       \bl{\ell}  = \frac{1}{\sqrt{2}}\left( \bl{e}_0 + \bl{e}_3\right) \;&, \;\;   \bl{n}  = \frac{1}{\sqrt{2}}\left( \bl{e}_0 - \bl{e}_3\right) \;, \\
       \bl{m}  = \frac{1}{\sqrt{2}}\left( \bl{e}_1 + i\,\bl{e}_2\right) \,&, \;\;
 \bl{\bar{m}}  = \frac{1}{\sqrt{2}}\left( \bl{e}_1 - i\,\bl{e}_2\right) \,.
     \end{array}
   \right.
\end{equation*}
In terms of this frame, the inverse metric is given by
\begin{equation*}
 g^{ab}\partial_{a}\partial_{b}  = -(\bl{\ell} \,\bl{n} + \bl{n} \,\bl{\ell}) +  (\bl{m} \,\bl{\bar{m}}  + \bl{\bar{m}} \, \bl{m})\,,
\end{equation*}
so that the only nonvanishing inner products between the vector fields of the null tetrad frame are
\begin{equation*}
  \ell^a\,n_a = -1  \quad \textrm{and} \quad   m^a\,\bar{m}_a = 1  \,.
\end{equation*}
Particularly, the vectors of the null tetrad frame are all lightlike. In terms of this basis, the Killing tensor \eqref{KillingT1} can be compactly written as
\begin{equation*}
\bl{K}  = -S_1(\bl{\ell} \,\bl{n} + \bl{n} \,\bl{\ell}) -S_2  (\bl{m} \, \bl{\bar{m}} + \bl{\bar{m}} \, \bl{m})\,.
\end{equation*}

Besides the pleasant algebraic properties of the null frame just introduced, it also has great geometric significance. Indeed, one can check that the following relations hold:
\begin{equation}\label{Shearfree}
  \left.
    \begin{array}{ll}
     d\boldsymbol{\ell}\wedge\boldsymbol{\ell}\wedge\boldsymbol{m} &  =0\quad\text{and }\quad d\boldsymbol{m}\wedge\boldsymbol{\ell}\wedge\boldsymbol{m}=0\,,\\
d\boldsymbol{\ell}\wedge\boldsymbol{\ell}\wedge \bl{\bar{m}} &  =0\quad\text{and }\quad d\bl{\bar{m}}\wedge\boldsymbol{\ell}\wedge\bl{\bar{m}}=0\,,\\
d\boldsymbol{n}\wedge\boldsymbol{n}\wedge\boldsymbol{m}  &  =0\quad\text{and }\quad d\boldsymbol{m} \wedge\boldsymbol{n}\wedge\boldsymbol{m} =0\,, \\
d\boldsymbol{n}\wedge\boldsymbol{n}\wedge\bl{\bar{m}} &  =0\quad\text{and }\quad d\bl{\bar{m}}\wedge\boldsymbol{n}\wedge\bl{\bar{m}} =0\,.
    \end{array}
  \right.
\end{equation}
Where in the latter equation the vector fields of the null tetrad should, actually, be understood as their correspondent 1-forms, with the map between vector and 1-forms being provided by the metric tensor, the so-called lowering of indices. The first row of Eq. \eqref{Shearfree}, along with the Frobenius theorem, implies that the surfaces spanned by the vector fields that are orthogonal to both $\bl{\ell}$ and $\bl{m}$ form a locally integrable foliation of the spacetime \cite{Frankel}. Since the vector fields $\bl{\ell}$ and $\bl{m}$ are both null and orthogonal to each other, it follows that these integrable surfaces are spanned by the vectors $\bl{\ell}$ and $\bl{m}$ themselves. Due to the fact that the maximum dimension of a null subspace in four dimensions is two, we say that $Span\{\boldsymbol{\ell},\boldsymbol{m}\}$ is a maximally isotropic integrable distribution. In the same fashion,  the three remaining equations in \eqref{Shearfree} imply that the maximally isotropic distributions  $Span\{\boldsymbol{\ell},\bl{\bar{m}}\}$, $Span\{\boldsymbol{n}%
,\boldsymbol{m} \}$ and $Span\{\boldsymbol{n},\bl{\bar{m}}\}$ are all integrable. An equivalent way to phrase this is saying that the
null vector fields $\boldsymbol{\ell}$ and $\boldsymbol{n}$ are geodesic and shear-free \cite{RobinsonManifolds}.

Another important feature of this null tetrad is that the real vector fields $\bl{\ell}$ and $\bl{n}$ are principal null directions of the spacetime, namely the Weyl scalars
\begin{equation*}
  \Psi_0 = C_{abcd}\ell^a m^b \ell^c m^d   \; \textrm{ and } \; \Psi_4 = C_{abcd}n^a \bar{m}^b n^c \bar{m}^d
\end{equation*}
are both vanishing, with $C_{abcd}$ standing for the Weyl tensor. Nevertheless, such null vector fields are not repeated principal null directions, inasmuch as the Weyl scalars
\begin{equation*}
  \Psi_1 = C_{abcd}\ell^a n^b \ell^c m^d   \; \textrm{ and } \; \Psi_3 = C_{abcd}n^a \ell^b n^c \bar{m}^d
\end{equation*}
are generally non-vanishing \cite{Bat1}. Although it is interesting to point out that $\Psi_1 = \Psi_3$, so that whenever $\bl{\ell}$ is a repeated principal null direction, so will be $\bl{n}$. Nevertheless, if we are interested in the subclass of the spacetimes \eqref{BFmetric} obeying Einstein's vacuum equation  $R_{ab}=\Lambda g_{ab}$, with $R_{ab}$ denoting the Ricci tensor and $\Lambda$ being a cosmological constant, then we can impose that $\Psi_1 = \Psi_3 = 0$ without loss of generality, as a consequence of the so-called Goldberg-Sachs theorem \cite{Goldberg-Sachs,GS-Conformal,Plebanski2,BatGS}. Indeed, the latter theorem states that, for vacuum spacetimes, a vector field points in a repeated principal null direction if, and only if, it is geodesic and shear-free, a condition obeyed by the null vectors $\bl{\ell}$ and $\bl{n}$. This fact has been fully exploited in Ref. \cite{AnabalonBatista} in order to solve Einstein's vacuum equation for the class of spacetimes \eqref{BFmetric}.

The aim of the present work is to broaden the results of Ref. \cite{AnabalonBatista} by allowing the existence of gauge fields instead of assuming vacuum. Notwithstanding, a drawback that we immediately face is that the Goldberg-Sachs theorem does not hold in such a general case. However, the same article that establishes Goldberg-Sachs theorem provides the result that if a real null direction $\bl{k}$ is geodesic and shear-free and the constraints
\begin{equation}\label{ConditionRab}
  R_{ab}k^ak^b =0\;,\;\; R_{ab}k^a \eta^b   =0\;,\;\;   R_{ab}\eta^a\eta^b = 0
\end{equation}
hold for some complex null vector $\bl{\eta}$ orthogonal to $\bl{k}$, then the vector field $\bl{k}$ is a repeated principal null direction \cite{Goldberg-Sachs}. As will be explained in the following section, it turns out that if we assume that the gauge field is aligned with the null directions $\bl{\ell}$ and $\bl{n}$ than the conditions \eqref{ConditionRab} hold and, as a consequence, we can assume that $\bl{\ell}$ and $\bl{n}$ are repeated principal null directions, which will provide a great simplification in our integration process.


\section{Using a Convenient Parametrization for the Metric}\label{Sec.Metric}

In this section we will make a change on the notation used to describe the metric (\ref{BFmetric}) that will be suitable for our purposes.
From the condition (\ref{detF}), it follows that $G_1^{\tau\sigma}$ is determined in terms of $G_1^{\tau\tau}$ and $G_1^{\sigma\sigma}$ and, analogously, $G_2^{\tau\sigma}$ is known once $G_2^{\tau\tau}$ and $G_2^{\sigma\sigma}$ are established. But, instead of using the functions $G_1^{\tau\tau}$ and $G_1^{\sigma\sigma}$ as our unknowns, we will use the functions $P_1$ and $N_1$ defined by
\begin{equation*}\label{G1}
  N_1(x)  = \frac{1}{ G_1^{\sigma\sigma}\Delta_1} \;\;,\;\; P_1(x) = - \,\sqrt{\frac{G_1^{\tau\tau}}{G_1^{\sigma\sigma}}}  \,.
\end{equation*}
Analogously, rather than using the functions $G_2^{\tau\tau}$ and $G_2^{\sigma\sigma}$ we shall use the functions $P_2$ and $N_2$ defined by
\begin{equation*}\label{G2}
  N_2(y)  = \frac{1}{ G_2^{\sigma\sigma}\Delta_2} \;\;,\;\; P_2(y) =  \sqrt{\frac{G_2^{\tau\tau}}{G_2^{\sigma\sigma}}}  \,.
\end{equation*}
With this new parametrization, we can check that the line element of our spacetime is given by:
\begin{align}
 d&s^{2}=(S_1+S_2)\Bigg[ \frac{-\,N_{2}\,\Delta_{2}}{(P_{1}+P_{2})^{2}}\, \left( d\tau + P_{1}\,d\sigma\right)^{2}  \nonumber \\
&+ \frac{N_{1}\,\Delta_{1}}{(P_{1}+P_{2})^{2}} \left(  d\tau - P_{2}\,d\sigma \right)^{2}  +   \frac{dx^{2}}{\Delta_{1}}
+ \frac{dy^{2}}{\Delta_{2}}   \Bigg] .\label{LineElement}
\end{align}
In accordance with the Ref. \cite{AnabalonBatista}, the integration of Einstein's equation will lead to three independent cases depending on whether the functions $P_1'$ and $P_2'$ are zero or not, where the primes represent derivations with respect to the coordinate on which a function depends (for instance, $P_1' = \frac{dP_1(x)}{dx}$). Therefore, in our integration process we shall tackle these three cases separately.


\section{The Alignment Condition of the Gauge Field and its Consequences}\label{Sec.AlignementCondition}

In the present work we are going to consider the so-called Einstein-Yang-Mills theory, which models a gauge field $\bl{\mathcal{A}}$ interacting with the gravitational field trough minimal coupling, so that the action of the system is given by:
\begin{equation}\label{Action}
   \int \sqrt{|g|}\left[ R - 2 \Lambda -  \frac{1}{2\lambda}\textrm{Tr}\left( \mathcal{F}^{ab} \mathcal{F}_{ab} \right)  \right] d\tau d\sigma dx dy\,,
\end{equation}
where $g$ stands for the determinant of the metric, $R$ is the Ricci scalar,
$\lambda$ is a coupling constant, and $\bl{\mathcal{F}}$ is the field strength, a Lie algebra-valued 2-form defined by
\begin{equation}\label{FdA}
  \bl{\mathcal{F}} \eq d\bl{\mathcal{A}} + \bl{\mathcal{A}}\wedge \bl{\mathcal{A}} \,.
\end{equation}
The trace in Eq. \eqref{Action} should be taken over the internal degrees of freedom stemming from the Lie algebra. The equations of motion associated to the above action are
\begin{align}
  &R_{ab} + \left(\Lambda - \frac{1}{2}R \right) g_{ab} = T_{ab} \,, \label{EqMotion11} \\
   & \mathcal{D}\star\bl{\mathcal{F}} \equiv  d\star\bl{\mathcal{F}}   +  [ \bl{\mathcal{A}}, \star\bl{\mathcal{F}}] = 0 \,, \label{EqMotion12}
\end{align}
where $\mathcal{D}$ stands for the gauge group covariant derivative,
$\bl{\star\mathcal{F}}$ denotes the Hodge dual of $\bl{\mathcal{F}}$, and $T_{ab}$ is the energy-momentum tensor of the gauge field, defined by
\begin{equation*}
 T_{ab} = \frac{1}{\lambda}\textrm{Tr}\left( \mathcal{F}_{ac}\mathcal{F}_b^{\ph{b}c} - \frac{1}{4}\,g_{ab}\,\mathcal{F}_{cd}\mathcal{F}^{cd}  \,  \right)\,.
\end{equation*}
Particularly, in the abelian case we can ignore the last term in the right hand side of Eq. \eqref{FdA} as well as the commutator in Eq. (\ref{EqMotion12}), as they are identically zero in such a case.

We say that a real bivector $H_{ab}=H_{[ab]}$ is aligned with a real null direction $\bl{k}$ whenever the condition
\begin{equation*}
  H_{ab} \, k^a \, \eta^b \eq 0
\end{equation*}
holds for some complex null vector $\bl{\eta}$ linearly independent and orthogonal to $\bl{k}$. In the Newman-Penrose formalism this means that $\bl{k}$ is a principal null direction of the bivector $\bl{H}$ \cite{Newman:1961qr}, a fact that is neatly understood through spinorial formalism \cite{Peeling-Penrose}. In what follows, it will be assumed that the gauge field inherits the geometric properties of the spacetime. In particular, this means that the field strength  $\bl{\mathcal{F}}$ is aligned with the principal null directions $\bl{\ell}$ and $\bl{n}$ of the Weyl tensor of our spacetime, namely
\begin{equation}\label{Align}
   \mathcal{F}_{ab} \, \ell^a \, m^b \eq 0  \;\;\; \textrm{and} \;\;\;  \mathcal{F}_{ab} \, n^a \, m^b \eq 0 \,.
\end{equation}
Particularly, since $\bl{\mathcal{F}}$ is assumed to have real components, the complex conjugate of the above relations imply that $\mathcal{F}_{ab}  \ell^a  \bar{m}^b$ and $\mathcal{F}_{ab}  n^a  \bar{m}^b$ are also vanishing.
In addition, the gauge field will be assumed to be invariant under the Lie transport along the Killing vector fields $\partial_\tau$ and $\partial_\sigma$, so that the Lie derivative of $\mathcal{A}$ along these vectors vanish:
\begin{equation}\label{LieDeriv}
  \mathcal{L}_{\partial_\tau} \mathcal{A} \eq 0  \;\;\; \textrm{and} \;\;\;  \mathcal{L}_{\partial_\sigma} \mathcal{A} \eq 0 \,.
\end{equation}
The latter condition means that the components of  $\mathcal{A}$ in the coordinate frame do not depend on the cyclic coordinates $\tau$ and  $\sigma$.

An interesting algebraic consequence of the alignment condition is that the following eight components of the energy-momentum tensor vanish:
\begin{equation}\label{TabAlign1}
  \left.
     \begin{array}{ll}
      T_{ab} \ell^a \ell^b  \hspace{-0.09cm}= \hspace{-0.09cm} T_{ab} n^a n^b \hspace{-0.09cm}= \hspace{-0.09cm} T_{ab} m^a m^b \hspace{-0.09cm}= \hspace{-0.09cm} T_{ab} \bar{m}^a \bar{m}^b = 0 ,\\
       T_{ab} \ell^a m^b  \hspace{-0.09cm}= \hspace{-0.09cm} T_{ab} \ell^a \bar{m}^b \hspace{-0.09cm}= \hspace{-0.09cm} T_{ab} n^a m^b \hspace{-0.09cm}= \hspace{-0.09cm} T_{ab} n^a \bar{m}^b = 0 ,
     \end{array}
   \right.
\end{equation}
a fact that can be easily checked by using Eq. (\ref{Align}). In addition, it follows that
\begin{equation}\label{TabAlign2}
  T_{ab} \,\left( \ell^a n^b -  m^a \bar{m}^b \right) = 0\,,
\end{equation}
where the latter relation stems from the fact that the trace of the energy-momentum tensor of the gauge field is zero, $T_{ab}g^{ab}=0$.
These relations provide a great simplification on the integration of the equation of motion \eqref{EqMotion11}. Indeed, defining the tensor
\begin{equation}\label{EqMotion2}
  E_{ab} = R_{ab} + \left(\Lambda - \frac{1}{2}R \right) g_{ab} - T_{ab} \,,
\end{equation}
it follows that Einstein's field equation is given by $E_{ab} = 0$. The latter condition comprise a total of ten equations, since $E_{ab}$ is symmetric. Using a null tetrad frame, these ten components can be separated in the nine equations
\begin{equation}\label{Nine}
  \left.
    \begin{array}{ll}
     E_{ab} \ell^a \ell^b \hspace{-0.09cm}= \hspace{-0.09cm} E_{ab} n^a n^b \hspace{-0.09cm}= \hspace{-0.09cm}  E_{ab} m^a m^b
\hspace{-0.09cm}= \hspace{-0.09cm} E_{ab} \bar{m}^a \bar{m}^b \hspace{-0.09cm}= \hspace{-0.05cm}  0, \\
     E_{ab} \ell^a m^b \hspace{-0.09cm}= \hspace{-0.09cm} E_{ab} \ell^a \bar{m}^b \hspace{-0.09cm}= \hspace{-0.09cm}
E_{ab} n^a m^b \hspace{-0.09cm}= \hspace{-0.09cm} E_{ab} n^a \bar{m}^b \hspace{-0.09cm}= \hspace{-0.05cm}  0, \\
   E_{ab} ( \ell^a n^b -  m^a \bar{m}^b ) = 0,  \\
    \end{array}
  \right.
\end{equation}
plus the equation
\begin{equation}\label{EinstenEq}
  E_{ab} ( \ell^a n^b +  m^a \bar{m}^b ) = 0 \,.
\end{equation}
As consequence of conditions \eqref{TabAlign1} and \eqref{TabAlign2}, it turns out that the nine equations \eqref{Nine} do not depend on the gauge field $\bl{\mathcal{A}}$ at all, these differential equations involve just the functions of the metric. In other words, such equations are exactly the same as the ones faced in the vacuum problem. Since, for the class of metrics studied here, Einstein's equation in vacuum have
 already been solved in Ref. \cite{AnabalonBatista}, we can skip the integration of Eqs. (\ref{Nine}) and borrow the results from the latter reference.  Thus, we are left with only one component of Einstein's equation to solve, namely Eq. \eqref{EinstenEq}. In addition, we shall integrate the gauge field equation \eqref{EqMotion12}. Nevertheless, it is worth pointing out that the latter equation is, in a sense, contained in Einstein's equation, due to the fact that the  energy-momentum tensor is divergence-free, $\nabla^aT_{ab}=0$, although we do not exploit this fact in the sequel. Thus, at the end of the day, we essentially need to solve the differential equation Eq. (\ref{EinstenEq}) along with the gauge field equation (\ref{EqMotion12}).

The general gauge field can be written as
\begin{equation}\label{Aform}
 \mathcal{A} = \mathcal{A}_\tau\, d\tau +   \mathcal{A}_\sigma  d\sigma +  \mathcal{A}_x  dx +  \mathcal{A}_y  dy \,,
\end{equation}
where $\mathcal{A}_a$ are elements of the Lie algebra of the gauge group depending just on the coordinates $x$ and $y$, as a consequence of the hypothesis (\ref{LieDeriv}). Now, due to the alignment condition (\ref{Align}), it follows that
\begin{equation}\label{Fxy}
 \mathcal{F}_{xy}  \propto  \mathcal{F}_{ab} \, (\ell^a - n^a)(m^b - \bar{m}^b) = 0\,.
\end{equation}
Then, since $\mathcal{A}_a$ depends just on the coordinates $x$ and $y$, it turns out that the $1$-form
\begin{equation*}
  \bl{\mathcal{\tilde{A}}} \equiv  \mathcal{A}_x  dx +  \mathcal{A}_y  dy
\end{equation*}
can be seen as a gauge field on the submanifold $\tilde{M}$ spanned by the coordinates $\{x,y\}$, namely $\tilde{M}$ is a leaf of constant $\tau$ and $\sigma$. Denoting by $\tilde{d}$ the exterior derivative on $\tilde{M}$, it turns out that Eq. (\ref{Fxy}) can be written as
\begin{equation*}
 \tilde{d} \bl{\mathcal{\tilde{A}}} + \bl{\mathcal{\tilde{A}}}\wedge \bl{\mathcal{\tilde{A}}} = 0 \,,
\end{equation*}
which means that the curvature of the gauge connection $\bl{\mathcal{\tilde{A}}}$ vanishes, so that $\bl{\mathcal{\tilde{A}}}$ is related trough a gauge transformation to the zero gauge field. In other words,  $\bl{\mathcal{\tilde{A}}}$ can be gauged away. Thus, Eq. (\ref{Fxy}), which stems from the alignment condition, guarantees that we can perform a gauge transformation in such a way to make both $\mathcal{A}_x$ and $\mathcal{A}_y$ equal to zero. Hence, in the sequel, we will implicitly use this gauge freedom to write our gauge field as
\begin{equation}\label{A-tau-sigma}
   \mathcal{A} = \mathcal{A}_\tau\, d\tau +   \mathcal{A}_\sigma  d\sigma\,,
\end{equation}
where $\mathcal{A}_\tau$ and $\mathcal{A}_\sigma$ are elements of the Lie algebra depending just on the coordinates $x$ and $y$.

Then, continuing the imposition of the alignment condition, we find that the general solution for the constraints
\begin{align*}
 &\mathcal{F}_{ab} \, (\ell^a - n^a)(m^b + \bar{m}^b) = 0 \,, \\
  &\mathcal{F}_{ab} \, (\ell^a + n^a)(m^b - \bar{m}^b) = 0 \,,
\end{align*}
is provided by
\begin{equation}\label{AtauAsigma}
  \mathcal{A}_\tau  = \frac{\mathcal{B}_1  + \mathcal{B}_2}{P_1+P_2} \quad \textrm{and} \quad
\mathcal{A}_\sigma  = \frac{ P_1 \mathcal{B}_2  - P_2 \mathcal{B}_1 }{P_1+P_2} \,,
\end{equation}
where $\mathcal{B}_1 =\mathcal{B}_1(x)$ is an arbitrary element of the Lie algebra depending just on the coordinate $x$ while $\mathcal{B}_2 =\mathcal{B}_2(y)$ is an arbitrary element of the Lie algebra depending just on $y$. Finally, imposing the last constraint arising from the alignment condition, namely,
\begin{equation}\label{Ftausigma}
  \mathcal{F}_{\tau\sigma} \propto \mathcal{F}_{ab} \, (\ell^a + n^a)(m^b + \bar{m}^b) = 0\,,
\end{equation}
we conclude that
\begin{equation}\label{AA0}
  \mathcal{A}\wedge \mathcal{A}  = 0 \,,
\end{equation}
which means that the elements $\mathcal{B}_1$ and $\mathcal{B}_2$ should commute with each other.
Note that Eq. (\ref{AA0}) implies that the field strength $\bl{\mathcal{F}}$ is linear in the gauge field, just as in the abelian case.

Now, let us investigate the consequences of the gauge field equation of motion. Taking the Hodge dual of Eq. (\ref{EqMotion12}), we obtain that the gauge field equation is given by
\begin{equation}\label{EqmotionA2}
  \nabla^a \mathcal{F}_{ab} \ma [\mathcal{A}^a, \mathcal{F}_{ab}] = 0 \,.
\end{equation}
Then, due to the condition \eqref{Fxy}, it follows that the components $\nabla^a \mathcal{F}_{ax}$ and $\nabla^a \mathcal{F}_{ay}$ of the divergence of the two-form $\bl{\mathcal{F}}$ vanish identically, so that Eq. (\ref{EqmotionA2}) implies that
\begin{equation*}
  [\mathcal{A}^a, \mathcal{F}_{ax}] =  [\mathcal{A}^a, \mathcal{F}_{ay}] = 0 \,.
\end{equation*}
On the other hand, due to the fact that $\mathcal{F}_{\tau\sigma}$ is zero, as shown in Eq. (\ref{Ftausigma}), it follows that the contractions
$\mathcal{A}^a \mathcal{F}_{a\tau}$, $\mathcal{F}_{a\tau}\mathcal{A}^a $, $\mathcal{A}^a \mathcal{F}_{a\sigma}$ and   $\mathcal{F}_{a\sigma} \mathcal{A}^a$ are all vanishing, so that the field equation (\ref{EqmotionA2}) yields
\begin{equation*}
  \nabla^a \mathcal{F}_{a\tau} = \nabla^a \mathcal{F}_{a\sigma} = 0 \,.
\end{equation*}
Thus, it turns out that the single field equation (\ref{EqmotionA2}), along with the consequences of the alignment condition, implies that the following two equations should hold
\begin{align}
  &\nabla^a \mathcal{F}_{ab}  = 0 \,,\label{EqmotionA31}\\
  &[\mathcal{A}^a, \mathcal{F}_{ab}] = 0 \,. \label{EqmotionA32}
\end{align}
The equation (\ref{EqmotionA31}) is just the equation of motion of the abelian case. Moreover, recall that the condition \eqref{AA0} entails that the field strength has the same form of the one in the abelian case, $\bl{\mathcal{F}} = d\mathcal{A}$. Hence, we can conclude that our task of finding the solution for the Einstein-Yang-Mills system of equations, for arbitrary gauge group, is equivalent to solving the abelian problem supplied with the algebraic conditions \eqref{AA0} and \eqref{EqmotionA32}. This will be the path taken in the sequel: we will solve Einstein-Yang-Mills equations for the abelian case, the so-called Einstein-Maxwell system of equations, and then, in a posterior section, we will take these solutions and impose the two algebraic constraints  \eqref{AA0} and \eqref{EqmotionA32}, lifting the abelian solution to the case of a general gauge group. At this point, it is worth pointing out that although we are using the abelian solutions, which are simpler to obtain,  to construct non-abelian solutions, we should not interpret the latter as elementary generalizations of the former, as our non-abelian solutions can have a rich algebraic structure with no analogue in the abelian theory. Indeed, even in the simplest case in which the non-abelian gauge field is proportional to the $U(1)$ electromagnetic field, the non-abelian solution is physically distinguishable from the electromagnetic one, as argued in Ref. \cite{Yasskin:1975ag}.

\section{Solving the Equations of Motion for the Abelian Case}\label{Sec.Abelian Case}

As acknowledged previously, we will separate the integration process in three cases, depending on wether the derivatives $P_1'$ and $P_2'$ are zero or not. In the three cases, the procedure of integration will follow the same general steps: \textbf{(I)} Integrating the nine components of Einstein's equation that do not depend on the gauge field, Eqs. (\ref{Nine}), we will find the general form of the functions appearing on the metric, namely $S_1$, $S_2$, $N_1$, $N_2$, $\Delta_1$ and $\Delta_2$. Since these equations do not depend on the gauge field at all, as explained in the previous section, we can take advantage of the results obtained in Ref. \cite{AnabalonBatista} for the vacuum case. \textbf{(II)} Using the gauge field given in Eqs. (\ref{A-tau-sigma}) and (\ref{AtauAsigma}), and imposing the abelian gauge field equation, $\nabla^a \mathcal{F}_{ab}  = 0$, we will be left with the general form for $\mathcal{B}_1$ and $\mathcal{B}_2$, which are elements of the Lie algebra. \textbf{(III)} Imposing the remaining component of Einstein's field equation, Eq. (\ref{EinstenEq}), we will see that a pair of integration constants must be related to each other. After these three steps we will have accomplished our goal in this section, namely we will have found the most general solution for abelian Einstein-Yang-Mills system of equations in a background possessing two commuting Killing vectors, a nontrivial Killing tensor and a gauge field that is aligned with the principal null directions of the spacetime.



\subsection{Case $P1'\neq 0$ and $P2'\neq 0$}\label{SubSec.geral}

If we assume that $P_1(x)$ and $P_2(y)$ are both nonconstant, we can always make a coordinate choice such that
\begin{equation*}\label{P1P2-1}
  P_1(x) = x^2  \quad \textrm{and} \quad  P_2(y) = y^2  \,.
\end{equation*}
Indeed, this is attained by making a coordinate transformation $(x,y)\rightarrow (\hat{x} = \sqrt{P_1},\hat{y}= \sqrt{P_2})$ in the line element (\ref{LineElement}), along with the redefinition of the functions $N_1$, $\Delta_1$, $N_2$, $\Delta_2$ and then dropping the hats in the new coordinates.
For this choice of coordinates, by following steps completely analogous to the ones taken in Ref. \cite{AnabalonBatista}, one can see that the most general solution of the nine Eqs. (\ref{Nine}) is provided by
\begin{align*}
 N_1  = \frac{x^2}{(a_1 + b_1 x^2 ) (a_2 + b_2 x^2 )} ,\;  S_1  = \frac{a_3 + b_3 x^2 }{a_1 + b_1 x^2 }, \\
 N_2  = \frac{-\,y^2}{(a_1 - b_1 y^2 ) (a_2 - b_2 y^2 )} ,\;   S_2  = \frac{a_3 - b_3 y^2 }{ b_1 y^2 - a_1 } ,
\end{align*}
where $a$'s and $b$'s are arbitrary integration constants. Moreover, the functions $\Delta_1$ and $\Delta_2$ must be given by
\begin{align*}\label{Delta1-1}
  \Delta_1 = \frac{1}{x^2}\Bigg[ c_1 I_1^{5/2}J_1^{3/2} &+ c_2 I_1^{3}J_1  + c_3 I_1^{2}J_1^{2}  \\
 &+ \frac{\Lambda(a_1 b_3 - a_3 b_1)}{3b_1^2(a_2 b_1 - a_1 b_2)}   I_1 J_1 \Bigg]\,, \\
 \Delta_2 = \frac{1}{y^2}\Bigg[ c_4 I_2^{5/2}J_2^{3/2} &+ c_5 I_2^{3}J_2  - c_3 I_2^{2}J_2^{2}  \\
 &- \frac{\Lambda(a_1 b_3 - a_3 b_1)}{3b_1^2(a_2 b_1 - a_1 b_2)} I_2 J_2  \Bigg]\,,
\end{align*}
where the $c$'s are integration constants, which are arbitrary for the moment, whereas the $I$'s and $J$'s stand for the following functions:
\begin{equation}\label{I1J1}
  \left.
                                                        \begin{array}{ll}
                                                         I_1 = a_1 + b_1 x^2 \,,\;\;  J_1 = a_2 + b_2 x^2 \,,  \\
  I_2 = a_1 - b_1 y^2 \,,\;\;  J_2 = a_2 - b_2 y^2 \,.
                                                        \end{array}
                                                      \right.
\end{equation}

What remains to be integrated are Eqs. (\ref{EqmotionA31}) and (\ref{EinstenEq}). Assuming the form of the gauge field given by Eqs. (\ref{A-tau-sigma}) and (\ref{AtauAsigma}) and then imposing the gauge field equation $\nabla^a \mathcal{F}_{ab}  = 0$, we are led to the following general solution
\begin{equation*}\label{fhfhg}
  \left.
    \begin{array}{ll}
      \mathcal{B}_1 = \mathcal{Q}_1\, I_1^{1/2} \,J_1^{1/2} + \mathcal{Q}_3\, x^2 + \mathcal{Q}_4 \,,\\
      \mathcal{B}_2 =  \mathcal{Q}_2\, I_2^{1/2} \,J_2^{1/2} + \mathcal{Q}_3\, y^2 - \mathcal{Q}_4 \,,
    \end{array}
  \right.
\end{equation*}
where the $\mathcal{Q}$'s are arbitrary elements of the Lie algebra that are constant. Nevertheless, it turns out that the terms containing $\mathcal{Q}_3$ and $\mathcal{Q}_4$ are pure gauge, as they do not appear in none of the components of the field strength $\bl{\mathcal{F}}$. Therefore, without loss of generality, we can ignore these terms and simply write
\begin{equation}\label{B1B2-1}
  \left.
    \begin{array}{ll}
      \mathcal{B}_1 = \mathcal{Q}_1\, I_1^{1/2} \,J_1^{1/2} \,,\\
      \mathcal{B}_2 =  \mathcal{Q}_2\, I_2^{1/2} \,J_2^{1/2}  \,.
    \end{array}
  \right.
\end{equation}
The constant elements of the Lie algebra $\mathcal{Q}_1$ and $\mathcal{Q}_2$ should be interpreted as the charges generating the gauge field. Particularly, in the case of the electromagnetic gauge group, $U(1)$, these are the magnetic and electric charges respectively.

Finally, it remains to impose the last component of Einstein's equation, namely Eq. (\ref{EinstenEq}). This equation will be solved if, and only if, the integration constants $c_2$ and $c_5$, appearing in the functions $\Delta_1$ and $\Delta_2$, are related to each other by the equation
\begin{equation}\label{c5-geral}
  c_5 = \frac{(a_1 b_2 - a_2 b_1)}{2\lambda( a_3 b_1 - a_1 b_3 )} \textrm{Tr}\left(\mathcal{Q}_1\mathcal{Q}_1 - \mathcal{Q}_2\mathcal{Q}_2 \right) - c_2\,.
\end{equation}

The solution presented above is just a straightforward generalization of the charged Kerr-NUT-(A)dS class of spacetimes to the case where there are several electromagnetic fields decoupled from each other. Indeed, assuming that the gauge group is $U(1)$, a one-dimensional compact group, and performing the change of coordinates described in Appendix \ref{AppendixCoord}, we are left with the Kerr-Newman-NUT-(A)dS solution \cite{GrifPodol}.


\subsection{Case $P1'\neq 0$ and $P2'= 0$}\label{SubSec.Pleb}


Now, let us suppose that the function $P_1(x)$ is non-constant while $P_2(y)$ is a constant, henceforth denoted by $p_2$. Then, just as we did in the previous subsection, we can choose a coordinate system in which $x$ is some desired function of $P_1(x)$, without loss  of generality. Here, we will choose $x=(P_1 + p_2)^{-1}$, so that
\begin{equation*}
  P_1(x) = \frac{1}{x} - p_2  \quad \textrm{and} \quad  P_2(y) = p_2 \,.
\end{equation*}
In such a case, the most general solutions for Eqs. (\ref{Nine}) are given by
\begin{align*}
& N_1  = \frac{a_1}{4a_2\,x^2} \;,\;\;\;  S_1  = b\;, \; \Delta_1 = c_1 x^2 + c_2 x + c_3 \,,\\
& N_2  = \frac{a_1}{ (y^2 + a_2)^2} \;,\;\;   S_2  = y^2 + a_2 - b \;,\\
&\Delta_2 =-\frac{\Lambda}{3}\,y^4 -(c_1 + 2 a_2 \Lambda)y^2 + c_4 \,y + c_5  \,,
\end{align*}
where the parameters $a$'s, $b$ and $c$'s are, up to now, arbitrary integration constants.

Then, imposing the gauge field equation $\nabla^a \mathcal{F}_{ab}  = 0$ for the field given by Eqs. (\ref{A-tau-sigma}) and (\ref{AtauAsigma}), we find that the most general solution is provided by
\begin{equation*}\label{fhfhg2}
  \left.
    \begin{array}{ll}
      \mathcal{B}_1 = \mathcal{Q}_3\, \frac{1}{x} + \mathcal{Q}_4  \,,\\
      \mathcal{B}_2 =  \mathcal{Q}_1\,\frac{a_2-y^2}{a_2(a_2+y^2)}  + \mathcal{Q}_2\, \frac{2 y}{\sqrt{a_2}(a_2+y^2)}- \mathcal{Q}_4 \,,
    \end{array}
  \right.
\end{equation*}
where the $\mathcal{Q}$'s are arbitrary constant elements of the Lie algebra. Nevertheless, we can check that the components of the field strength $\bl{\mathcal{F}}$ do not depend on $\mathcal{Q}_3$ and $\mathcal{Q}_4$ at all, meaning that they can be eliminated from the gauge field $\mathcal{A}$ by a gauge transformation. Hence, we can, without loss of generality, assume that $\mathcal{Q}_3$ and $\mathcal{Q}_4$ are both zero, in which case we have
\begin{equation*}\label{B1B2-2}
  \left.
    \begin{array}{ll}
      \mathcal{B}_1 = 0  \,,\\
      \mathcal{B}_2 =  \mathcal{Q}_1\,\frac{a_2-y^2}{a_2(a_2+y^2)}  + \mathcal{Q}_2\, \frac{2 y}{\sqrt{a_2}(a_2+y^2)} \,.
    \end{array}
  \right.
\end{equation*}
The constant Lie algebra elements $\mathcal{Q}_1$ and $\mathcal{Q}_2$ are the charges that generate the gauge field.

Lastly, we need to impose the remaining component of Einstein's equation, Eq. (\ref{EinstenEq}). The latter equation is obeyed if, and only if, the constants $c_1$ and $c_5$ appearing on the expressions of $\Delta_1$ and $\Delta_2$ are related to each other by the following relation:
\begin{equation}\label{c5-Pleb}
  c_5 = a_2 c_1 + a_2^2 \Lambda + \frac{2}{a_1 a_2 \lambda}\,\textrm{Tr}\left( \mathcal{Q}_1 \mathcal{Q}_1 + \mathcal{Q}_2\mathcal{Q}_2 \right).
\end{equation}

The solution just presented is a generalization of a twisting but non-accelerating spacetime contained in the Pleba\'{n}ski-Demia\'{n}ski class of metrics which has, as a special case, the charged Taub-NUT-(A)dS spacetime \cite{GrifPodol}. Indeed, performing suitable coordinate transformations and choosing the gauge group to be the one of a single electromagnetic field, $U(1)$, we can write the latter solution in a known form, as described in Appendix \ref{AppendixCoord}. The fact that the acceleration parameter should be zero can be grasped from the established result that accelerated solutions generally do not have a nontrivial Killing tensor, instead they can possess a conformal Killing tensor \cite{Hugh_Killing,Stephani}.

Here we will not explicitly consider the case  $P1'= 0$ and $P2'\neq 0$, since this case can be obtained trivially from the one presented in this subsection. Indeed, by interchanging the coordinates $x$ and $y$ we can go from the case in which $P1'\neq 0$ and $P2'= 0$ to the case $P1'= 0$ and $P2'\neq 0$.

\subsection{Case $P1'= 0$ and $P2'= 0$}

Finally, we shall deal with the case in which both functions $P_1$ and $P_2$ are constants, henceforth denoted by
\begin{equation*}
  P_1(x) = p_1 \quad \textrm{and} \quad P_2(y) = p_2\,.
\end{equation*}
In such a case, we can easily perform a linear translation on the coordinates $\tau$ and $\sigma$ in order to make the line element diagonal. Indeed, defining the coordinates
\begin{equation*}
  t = \frac{1}{p_1+p_2}(\tau + p_1 \sigma) \quad \textrm{and}\quad  \phi = \frac{1}{p_1+p_2}(\tau - p_2 \sigma)\,,
\end{equation*}
and choosing the coordinate $x$ to be such that $\Delta_1(x)=1$, it follows that the line element (\ref{LineElement}) is given by
\begin{equation}\label{Metric-s1s2}
  ds^{2}=(S_1+S_2)\Bigg[   N_{1}\,d\phi^{2} - N_{2} \Delta_{2}    dt^{2}
 +   dx^{2}
+ \frac{dy^{2}}{\Delta_{2}}   \Bigg]  .
\end{equation}
In this new coordinate system, the null tetrad is given by
\begin{equation*}
  \left\{
     \begin{array}{ll}
       \bl{\ell}  &= \frac{1}{\sqrt{2(S_1+S_2)}}\left( \frac{1}{\sqrt{N_2 \Delta_2}} \partial_t + \sqrt{ \Delta_2}\partial_y  \right) \,,\\
   \bl{n}  &= \frac{1}{\sqrt{2(S_1+S_2)}}\left( \frac{1}{\sqrt{N_2 \Delta_2}} \partial_t - \sqrt{ \Delta_2}\partial_y  \right) \,, \\
       \bl{m}  &= \frac{1}{\sqrt{2(S_1+S_2)}}\left( \frac{-1}{\sqrt{N_1}} \partial_\phi + i\, \partial_x  \right) \,,\\
 \bl{\bar{m}}  &= \frac{1}{\sqrt{2(S_1+S_2)}}\left( \frac{-1}{\sqrt{N_1}} \partial_\phi - i\, \partial_x  \right) \,,
     \end{array}
   \right.
\end{equation*}
while the aligned gauge field is written as
\begin{equation*}
  \bl{\mathcal{A}} = \mathcal{B}_2  \,dt + \mathcal{B}_1  \, d\phi \,,
\end{equation*}
where we should recall that $\mathcal{B}_1$ and $\mathcal{B}_2$ are, for the moment, arbitrary elements of the Lie algebra depending on $x$ and $y$ respectively.

The integration process of the case considered here will be split into two distinct cases. Indeed, the components
$E_{ab}\ell^a m^b$, $E_{ab}n^a m^b$, and their complex conjugates will vanish, as they should, if, and only if,
\begin{equation*}
  S_1'\,S_2' = 0\,.
\end{equation*}
Thus, we have two possibilities: either one of the functions $S_1$ and $S_2$ is constant, while the other is not, or both functions $S_1$ and $S_2$ are constant. Since the solutions stemming from these two cases are quite different from each other, we will segregate the analysis of these two situations. In the first case we have two subcases, namely $S_1$ constant while $S_2$ is nonconstant and the opposite subcase in which $S_1$ is nonconstant while $S_2$ is constant. However, these two subcases are not intrinsically different from each other, since an interchange of the coordinates $x$ and $y$ would map one into the other, so that it is sufficient to tackle just one of the subcases, as we do in the sequel.

\subsubsection{The subcase $S1' = 0$ and $S2' \neq 0$}\label{SubSec.s1cte}

Now, let us suppose that $S_1$ is constant, henceforth denoted by $b$, while $S_2$ is nonconstant. In such a case we can always redefine the coordinate $y$, along with the functions $N_2$ and $\Delta_2$, in such a way that
\begin{equation*}
  S_1(x) = b \quad \textrm{and} \quad S_2(y) = y^2 - b \,.
\end{equation*}
For this choice, the general solution of the nine components of Einstein's equation that do not depend on the gauge field, Eqs. (\ref{Nine}), is given by
\begin{align*}
 &N_1 = a_1\,\sin^2(c_1 x + c_2) \;,\;\;\; N_2 = a_2\,y^{-4} \;,\\
&\Delta_2 = -\,\frac{\Lambda}{3} y^4 + c_1^2 y^2 + d_1 y + d_2 \,,
\end{align*}
where the parameters $a$'s, $c$'s and $d$'s are, for the time being, arbitrary integration constants. Then, imposing the gauge field equation $\nabla^a \mathcal{F}_{ab}  = 0$, we find that the most general solution for the components of the gauge field is provided by
\begin{equation*}
  \left.
    \begin{array}{ll}
      \mathcal{B}_1 = \mathcal{Q}_1\, \cos(c_1 x + c_2 ) + \mathcal{Q}_3  \,,\\
      \mathcal{B}_2 =  \mathcal{Q}_2\, \frac{1}{y}+ \mathcal{Q}_4 \,,
    \end{array}
  \right.
\end{equation*}
where the $\mathcal{Q}$'s are arbitrary constant elements of the Lie algebra. Nevertheless, it follows that $\mathcal{Q}_3$ and $\mathcal{Q}_4$ do not appear on the field strength $\bl{\mathcal{F}}$, so that they can be eliminated from the gauge field $\mathcal{A}$ by a gauge transformation. Thus, we can assume that $\mathcal{Q}_3$ and $\mathcal{Q}_4$ are both zero, in which case we have
\begin{equation*}
  \mathcal{B}_1 = \mathcal{Q}_1\, \cos(c_1 x + c_2 )   \quad \textrm{and} \quad  \mathcal{B}_2 =  \mathcal{Q}_2\, \frac{1}{y}\,.
\end{equation*}
Finally, imposing the remaining component of Einstein's equation, Eq. (\ref{EinstenEq}), we find that the constant $d_2$ must be related to the other integration constants by the following equation
\begin{equation}\label{d2-s1cte}
  d_2 = \frac{c_1^2}{2 a_1 \lambda}\,\textrm{Tr}(\mathcal{Q}_1\mathcal{Q}_1) \,+\,
\frac{1}{2 a_2 \lambda}\,\textrm{Tr}(\mathcal{Q}_2\mathcal{Q}_2)\,.
\end{equation}

The latter solution is a simple generalization of the Reissner-Nordstr\"{o}m metric to the case where there are several decoupled electromagnetic fields, as explicitly shown in Appendix \ref{AppendixCoord}. See Ref. \cite{ReissnerNord} for an account on the Reissner-Nordstr\"{o}m spacetimes in the presence of a cosmological constant.

\subsubsection{The subcase $S1' =  S2' =0$}\label{SubSec.s1s2cte}

Now, let us suppose that $S_1$ and $S_2$ are both constant, so that we can write
\begin{equation*}
  S_1(x) = b_1 \quad \textrm{and} \quad S_2(y) = b_2 - b_1 \,,
\end{equation*}
where $b_1$ and $b_2$ are arbitrary constants. In this case, it is useful to redefine the coordinate $y$, along with the function $N_2$, so that the relation
\begin{equation*}
   \Delta_2(y) = 1
\end{equation*}
holds, which represents no loss of generality. For this choice, the general solution of the nine components of Einstein's equation that do not depend on the gauge field, Eqs. (\ref{Nine}), is provided by
\begin{align*}
& N_1 = a_1 \sin^2(c_1 x + c_2)\,, \\
&  N_2 = a_1 a_2 \sin^2\left(y\sqrt{2 b_2 \Lambda - c_1^2} + c_3\right) \,,
\end{align*}
where the $a$'s and $c$'s are, up to this moment, arbitrary integration constants. Next, demanding the gauge field equation $\nabla^a \mathcal{F}_{ab}  = 0$ to be obeyed, we find out that the most general solution is provided by
\begin{equation*}
  \left.
    \begin{array}{ll}
      \mathcal{B}_1 = \mathcal{Q}_1\, \cos(c_1 x + c_2 ) + \mathcal{Q}_3  \,,\\
      \mathcal{B}_2 =  \mathcal{Q}_2\, \cos\left(y\sqrt{2 b_2 \Lambda - c_1^2} + c_3\right) + \mathcal{Q}_4 \,,
    \end{array}
  \right.
\end{equation*}
where the $\mathcal{Q}$'s are arbitrary constant elements of the Lie algebra. However, it turns out that the field strength $\bl{\mathcal{F}}$ do not depend on $\mathcal{Q}_3$ and $\mathcal{Q}_4$, which means that they can be eliminated from the gauge field $\mathcal{A}$ by a gauge transformation. Thus, without loss of generality, we can assume that $\mathcal{Q}_3$ and $\mathcal{Q}_4$ are both zero, in which case we have
\begin{equation*}
  \left.
    \begin{array}{ll}
      \mathcal{B}_1 = \mathcal{Q}_1\, \cos(c_1 x + c_2 )   \,,\\
      \mathcal{B}_2 =  \mathcal{Q}_2\, \cos\left(y\sqrt{2 b_2 \Lambda - c_1^2} + c_3\right)\,.
    \end{array}
  \right.
\end{equation*}
Lastly, imposing the remaining component of Einstein's equation, Eq. (\ref{EinstenEq}), we are led to the fact that the constant $a_1$ must be given in terms of the other integration constants by
\begin{equation}\label{d2-s1s2cte}
  a_1 = \frac{ a_2 c_1^2\,\textrm{Tr}(\mathcal{Q}_1\mathcal{Q}_1) + ( 2b_2\Lambda- c_1^2) \textrm{Tr}(\mathcal{Q}_2\mathcal{Q}_2) }{2 a_2 b_2
( c_1^2 - b_2\Lambda) \lambda} \,.
\end{equation}

The spacetime just presented is the direct product of two spaces of constant curvature, one being a two-dimensional (anti-)de Sitter space and the other being the two-dimensional sphere. Thus, the above solution is a generalization of the charged Nariai spacetime \cite{BatNariai}, as explicitly shown in Appendix \ref{AppendixCoord}.


\section{General Gauge Group}\label{Sec.General Gauge}

The solutions presented in the previous section are valid for any abelian gauge group. In themselves, these solutions do not represent any novelty, as they are all simple generalizations of well-known $U(1)$ charged solutions to the case where there are several electromagnetic fields decoupled from each other. Nevertheless, as a whole, the integration performed above has a valuable practical significance, inasmuch as we have exhausted all solutions of Einstein-Yang-Mills theory for abelian groups in a background possessing two commuting Killing vectors and one nontrivial Killing tensor when the gauge field is aligned with the principal null directions of the spacetime. Thus, now we know that no new solutions can be found inside the entire class of spacetimes considered here. Now, the aim of the present section is to extend the integration process to the case of an arbitrary gauge group. This, in contrast to the abelian case, will lead to solutions that, as far as the authors know, have not been described in the literature yet.

As discussed in section \ref{Sec.AlignementCondition}, the general solution for an arbitrary gauge group can be found from the solutions of the abelian problem. For this purpose, all we need to do is to impose the algebraic conditions (\ref{AA0}) and (\ref{EqmotionA32}) to the solutions already found in the previous section. Thus, we shall take the latter solutions and impose the algebraic constraints
\begin{equation*}
   [\mathcal{A}_\tau, \mathcal{A}_\sigma] = 0 \;, \;\;\; [\mathcal{A}^a, \mathcal{F}_{ax}] = 0
\;\; \textrm{and} \;\; [\mathcal{A}^a, \mathcal{F}_{ay}] = 0 \,,
\end{equation*}
which are equivalent to the conditions (\ref{AA0}) and (\ref{EqmotionA32}).
Nicely, for all the cases considered in the latter section, these three constraints boils down to a single condition, namely the charges $\mathcal{Q}_1$ and $\mathcal{Q}_2$ must commute,
\begin{equation}\label{Q1Q2}
  [\mathcal{Q}_1, \mathcal{Q}_2] = 0 \,.
\end{equation}
So, for any gauge group, the most general solution for the class of spacetimes considered here are the solutions obtained in the previous section supplemented by the condition (\ref{Q1Q2}). Particularly, in the abelian case the condition (\ref{Q1Q2}) represents no constraint at all, since in such a case all elements of the Lie algebra commute with each other, so that $\mathcal{Q}_1$ and $\mathcal{Q}_2$ can be completely arbitrary elements. Thus, if  $k$ is the dimension of the gauge group, the number of charge parameters in the abelian case is $2k$, inasmuch as a general element of the Lie algebra has $k$ independent components.

An immediate solution for the constraint (\ref{Q1Q2}) that is valid for any gauge group is to choose $\mathcal{Q}_1$ arbitrarily and assume that $\mathcal{Q}_2$ to be proportional to $\mathcal{Q}_1$,
\begin{equation}\label{Q1-eQ2}
  \mathcal{Q}_2 = e\, \mathcal{Q}_1 \,,
\end{equation}
where $e$ is some arbitrary real constant. Indeed, if Eq. (\ref{Q1-eQ2}) holds then $\mathcal{Q}_1$ and $\mathcal{Q}_2$ trivially commute with each other. In such a case the number of charge parameters is equal to $k+1$, with $k$ standing for the dimension of the gauge group, since $\mathcal{Q}_1$ have $k$ components and $e$ is another independent parameter. This particular type of solution have already been described in Ref. \cite{Yasskin:1975ag}. Notwithstanding, although (\ref{Q1-eQ2}) represents a solution for Eq. (\ref{Q1Q2}), generally it is not the most general one. However, the most general solution of Eq. (\ref{Q1Q2}) cannot be  explicitly worked out without declaring the gauge group of interest. Thus, with the aim of being more explicit, we work out the solutions for two non-abelian gauge groups, $SU(2)$ and $SO(4)$.

\subsection{Group $SU(2)$}

In this subsection, let us suppose that the gauge group is $SU(2)$ and denote a basis for its Lie algebra by $\{\mathcal{L}_\alpha\}$. In order to be explicit, we shall introduce the following representation:
\begin{equation*}
  \mathcal{L}_1 \hspace{-0.1cm}= \hspace{-0.1cm} \frac{-i}{2}\left[
                     \begin{array}{cc}
                       0 & 1 \\
                       1 & 0 \\
                     \end{array}
                   \right] ,\,
 \mathcal{L}_2 \hspace{-0.1cm}= \hspace{-0.1cm} \frac{-i}{2}\left[
                     \begin{array}{cc}
                       0 & -i \\
                       i & 0 \\
                     \end{array}
                   \right] ,\,
 \mathcal{L}_3 \hspace{-0.1cm}= \hspace{-0.1cm} \frac{-i}{2}\left[
                     \begin{array}{cc}
                       1 & 0 \\
                       0 & -1 \\
                     \end{array}
                   \right] ,
\end{equation*}
whose algebra is given by
\begin{equation*}
  [\mathcal{L}_1, \mathcal{L}_2 ] = \mathcal{L}_3 \,,\;  [\mathcal{L}_2, \mathcal{L}_3 ] = \mathcal{L}_1 \,,\;
[\mathcal{L}_3, \mathcal{L}_1 ] = \mathcal{L}_2 \,,
\end{equation*}
which can be summarized by the following relation:
\begin{equation*}
  [\mathcal{L}_\alpha, \mathcal{L}_\beta ] = \varepsilon_{\alpha\beta}^{\phantom{\alpha\beta}\gamma}\,\mathcal{L}_\gamma \,,
\end{equation*}
where $\varepsilon_{\alpha\beta\gamma}$ is the three-dimensional Levi-Civita symbol. A metric in the Lie algebra vector space is then given by
\begin{equation*}
  \langle \mathcal{L}_\alpha, \mathcal{L}_\beta \rangle = \textrm{Tr}\left( \mathcal{L}_\alpha  \mathcal{L}_\beta\right) =
-\frac{1}{2}\,\delta_{\alpha\beta} \,.
\end{equation*}

Since the charges $\mathcal{Q}_1$ and $\mathcal{Q}_2$ are, in principle, arbitrary elements of the Lie algebra they are given by
\begin{align*}
  &\mathcal{Q}_1 = q_1^{\;\alpha} \mathcal{L}_\alpha =  q_1^{\;1} \mathcal{L}_1 +  q_1^{\;2} \mathcal{L}_2 + q_1^{\;3} \mathcal{L}_3 \,,\\
  &\mathcal{Q}_2 = q_2^{\;\alpha} \mathcal{L}_\alpha =  q_2^{\;1} \mathcal{L}_1 +  q_2^{\;2} \mathcal{L}_2 + q_2^{\;3} \mathcal{L}_3 \,,
\end{align*}
where the six ``charges'' $q_1^{\;\alpha}$ and $q_2^{\;\alpha}$ are, for the moment, arbitrary.
Then, the constraint (\ref{Q1Q2}), namely the requirement that  $\mathcal{Q}_1$ and $\mathcal{Q}_2$ commute, is transcribed as
\begin{equation}\label{qq}
  q_1^{\;\alpha}\,  q_2^{\;\beta} \, \varepsilon_{\alpha\beta\gamma} = 0 \,,
\end{equation}
whose most general solution is that the ``three-vectors'' $q_1^\alpha$ and $q_2^\alpha$ are proportional to each other,
\begin{equation}\label{q1-eq2}
  q_2^{\;\alpha} = e\,q_1^{\;\alpha} \,,
\end{equation}
where $e$ is an arbitrary parameter. This $SU(2)$ solution have already been described in Ref. \cite{Yasskin:1975ag}. Finally, note that the traces that appear in the solutions of Sec. \ref{Sec.Abelian Case} are written as
\begin{align*}
 & \textrm{Tr}\left( \mathcal{Q}_1  \mathcal{Q}_1\right) = -\frac{1}{2}\left[ (q_1^1)^2 + (q_1^2)^2 + (q_1^3)^2 \right] \, ,\\
& \textrm{Tr}\left( \mathcal{Q}_2  \mathcal{Q}_2\right) =  -\frac{e^2}{2}\left[ (q_1^1)^2 + (q_1^2)^2 + (q_1^3)^2 \right] \,.
\end{align*}
Eq. (\ref{q1-eq2}) reveals that for the group $SU(2)$ the most general solution is provided by the simple solution (\ref{Q1-eQ2}). In the next subsection we will investigate the gauge group $SO(4)$ and explicitly show that, depending on the gauge group, the moduli space can be richer, with other solutions besides the simple one shown in Eq. (\ref{Q1-eQ2}).

\subsection{Group $SO(4)$}\label{Sec.SO4}

Now, let us assume that the gauge group is $SO(4)$, whose Lie algebra can be generated by the antisymmetric $4\times 4$ matrices
\begin{align*}
   \mathcal{M}_{12} = \frac{1}{2}\left[
                        \begin{array}{cccc}
                          0 & 1 & 0 & 0 \\
                          -1 & 0 & 0 & 0 \\
                          0 & 0 & 0 & 0 \\
                          0 & 0 & 0 & 0 \\
                        \end{array}
                      \right] ,\;
\mathcal{M}_{13} =  \frac{1}{2}\left[
                        \begin{array}{cccc}
                          0 & 0 & 1 & 0 \\
                          0 & 0 & 0 & 0 \\
                          -1 & 0 & 0 & 0 \\
                          0 & 0 & 0 & 0 \\
                        \end{array}
                      \right]\,, \\
 \mathcal{M}_{14} =  \frac{1}{2}\left[
                        \begin{array}{cccc}
                          0 & 0 & 0 & 1 \\
                          0 & 0 & 0 & 0 \\
                          0 & 0 & 0 & 0 \\
                          -1 & 0 & 0 & 0 \\
                        \end{array}
                      \right] ,\;
\mathcal{M}_{23} =  \frac{1}{2}\left[
                        \begin{array}{cccc}
                          0 & 0 & 0 & 0 \\
                          0 & 0 & 1 & 0 \\
                          0 & -1 & 0 & 0 \\
                          0 & 0 & 0 & 0 \\
                        \end{array}
                      \right]\,, \\
 \mathcal{M}_{24} =  \frac{1}{2}\left[
                        \begin{array}{cccc}
                          0 & 0 & 0 & 0 \\
                          0 & 0 & 0 & 1 \\
                          0 & 0 & 0 & 0 \\
                          0 & -1 & 0 & 0 \\
                        \end{array}
                      \right] ,\;
\mathcal{M}_{34} =  \frac{1}{2}\left[
                        \begin{array}{cccc}
                          0 & 0 & 0 & 0 \\
                          0 & 0 & 0 & 0 \\
                          0 & 0 & 0 & 1 \\
                          0 & 0 & -1 & 0 \\
                        \end{array}
                      \right]\,.
\end{align*}
Instead of using the above basis of generators, it is neater to adopt the basis $\{\mathcal{L}_\alpha , \widetilde{\mathcal{L}}_\alpha\}$, defined by
\begin{align*}
 \mathcal{L}_1 =  \mathcal{M}_{12} +  \mathcal{M}_{34}\; &,\;\;   \mathcal{L}_2 =  \mathcal{M}_{14} +  \mathcal{M}_{23} \; ,\\
\mathcal{L}_3 =  \mathcal{M}_{13} & -  \mathcal{M}_{24} ,\; \\
\widetilde{\mathcal{L}}_1 =  \mathcal{M}_{12} -  \mathcal{M}_{34}\; &,\;\;   \widetilde{\mathcal{L}}_2 =  \mathcal{M}_{23} -  \mathcal{M}_{14} \; ,\\
\widetilde{\mathcal{L}}_3 =  \mathcal{M}_{13} & + \mathcal{M}_{24} .
\end{align*}
The nice thing about this basis is that the algebra obeyed by its elements consists of two independent copies of $SU(2)$ Lie algebra. More precisely, one can check that the following algebra holds
\begin{align*}
  [\mathcal{L}_\alpha, \mathcal{L}_\beta ] = \varepsilon_{\alpha\beta}^{\phantom{\alpha\beta}\gamma}\,\mathcal{L}_\gamma \; & , \;\;
[\widetilde{\mathcal{L}}_\alpha, \widetilde{\mathcal{L}}_\beta ] = \varepsilon_{\alpha\beta}^{\phantom{\alpha\beta}\gamma}\,\widetilde{\mathcal{L}}_\gamma \;,\\
 [\mathcal{L}_\alpha & ,  \widetilde{\mathcal{L}}_\beta ] = 0\,.
\end{align*}
In terms of this basis, a symmetric bilinear form in the Lie algebra vector space is then provided by
\begin{align*}
 & \langle \mathcal{L}_\alpha, \mathcal{L}_\beta \rangle = \textrm{Tr}\left( \mathcal{L}_\alpha  \mathcal{L}_\beta\right) =
- \delta_{\alpha\beta} \, ,  \\
& \langle \widetilde{\mathcal{L}}_\alpha, \widetilde{\mathcal{L}}_\beta \rangle =
\textrm{Tr}\left( \widetilde{\mathcal{L}}_\alpha  \widetilde{\mathcal{L}}_\beta\right) =
- \delta_{\alpha\beta} \, ,  \\
& \langle \mathcal{L}_\alpha, \widetilde{\mathcal{L}}_\beta \rangle = \textrm{Tr}\left( \mathcal{L}_\alpha  \widetilde{\mathcal{L}}_\beta\right) =
0\, .  \\
\end{align*}
In this basis, the general elements $\mathcal{Q}_1$ and $\mathcal{Q}_2$ can be written as
\begin{align*}
  &\mathcal{Q}_1 = q_1^{\;\alpha} \mathcal{L}_\alpha +  \tilde{q}_1^{\;\alpha} \widetilde{\mathcal{L}}_\alpha  \,,\\
  &\mathcal{Q}_2 = q_2^{\;\alpha} \mathcal{L}_\alpha +  \tilde{q}_2^{\;\alpha} \widetilde{\mathcal{L}}_\alpha \,,
\end{align*}
where the twelve ``charges'' $q_1^{\;\alpha}$, $\tilde{q}_1^{\;\alpha}$, $q_2^{\;\alpha}$ and $\tilde{q}_2^{\;\alpha}$ are, for the moment, arbitrary. Then, imposing that  $\mathcal{Q}_1$ and $\mathcal{Q}_2$ commute with each other lead us to the equation
\begin{equation*}
   q_1^{\;\alpha}\,  q_2^{\;\beta} \, \varepsilon_{\alpha\beta}^{\phantom{\alpha\beta}\gamma} \mathcal{L}_\gamma +
\tilde{q}_1^{\;\alpha}\,  \tilde{q}_2^{\;\beta} \, \varepsilon_{\alpha\beta}^{\phantom{\alpha\beta}\gamma} \widetilde{\mathcal{L}}_\gamma  = 0 \,,
\end{equation*}
whose most general solution is
\begin{equation}\label{q1-eq2-2}
  q_2^{\;\alpha} = e\,q_1^{\;\alpha} \quad \textrm{and} \quad  \tilde{q}_2^{\;\alpha} = \tilde{e}\,\tilde{q}_1^{\;\alpha} \,,
\end{equation}
where $e$ and $\tilde{e}$ are arbitrary real parameters independent from each other. Particularly, note that the traces that appear in the solutions of Sec. \ref{Sec.Abelian Case} can be written as
\begin{align*}
 & \textrm{Tr}\left( \mathcal{Q}_1  \mathcal{Q}_1\right) = - \sum_{\alpha=1}^3 (q_1^{\;\alpha})^2 +  (\tilde{q}_1^{\;\alpha})^2\, ,\\
& \textrm{Tr}\left( \mathcal{Q}_2  \mathcal{Q}_2\right) =  - \sum_{\alpha=1}^3 e^2(q_1^{\;\alpha})^2 +  \tilde{e}^2(\tilde{q}_1^{\;\alpha})^2 \,.
\end{align*}
Since generally we have $e\neq \tilde{e}$  it follows that $\mathcal{Q}_1$ is not proportional to $\mathcal{Q}_2$, so that the general solution (\ref{q1-eq2-2}) is not of the simple type presented in Eq. (\ref{Q1-eQ2}). Thus, the most general solution for the group $SO(4)$ has eight charge parameters, six stemming from the components of $\mathcal{Q}_1$ plus the two parameters $e$ and $\tilde{e}$. To the best of authors' knowledge, this solution has not been described in the literature yet. In particular, it is not included in the class of solutions presented in Ref. \cite{Yasskin:1975ag}, which is contained in the simple case  $\mathcal{Q}_2 = e\mathcal{Q}_1$.

\section{Conclusions}\label{Sec.Conclusions}

In this work we addressed the problem of analytically solving Einstein-Yang-Mills equations for the class of spacetimes possessing two commuting Killing vectors and one nontrivial Killing tensor for the case in which a natural null tetrad is defined, namely condition (\ref{detF}) holds. We have supposed that the gauge field inherits the geometric properties of the spacetime, so that it is aligned with the principal null directions of the Weyl tensor and its components are independent of the cyclic coordinates. Within this class of solutions, we have been able to perform the full integration for an arbitrary gauge group, so that now we can state that no new solution can be found inside the broad class considered here. In particular, when the gauge group is $U(1)$, this class contains the charged Kerr-NUT-(A)dS spacetime. Notwithstanding, solutions that have not been described in the literature yet have also been found, as explicitly exemplified by the general solution for the gauge group $SO(4)$, worked out in Sec. \ref{Sec.SO4}.

\begin{acknowledgments}
C. B. would like to thank Conselho Nacional de Desenvolvimento Cient\'{\i}fico e Tecnol\'ogico (CNPq) for the partial financial support and to Universidade Federal de Pernambuco for the support through the Qualis A reward.
\end{acknowledgments}

\appendix

\section{ Coordinate Transformations}\label{AppendixCoord}

As it has been pointed out in the previous sections, for the case of the abelian gauge group $U(1)$, the gauge group of electromagnetism, the solutions obtained here have already been described in the literature. The aim of the present appendix is to show explicitly the coordinate transformations that take our solutions to its known forms.  In order to accomplish this goal, we shall deal with each three cases considered above separately.


\subsection{Case $P'_1\neq0$ and $P'_2\neq0$}

Assuming that the gauge group is $U(1)$, let us consider the solution presented in section \ref{SubSec.geral}, for the case $P_1'\neq0$ and $P_2'\neq0$, and prove that this solution is actually the Kerr-Newman-NUT-(A)dS spacetime. To show this, we start from the metric (\ref{LineElement}),
with the functions $P_1$, $P_2$, $S_1$, $S_2$, $N_1$, $N_2$, $\Delta_1$ and $\Delta_2$ given by the ones shown in Sec. \ref{SubSec.geral}. Then, instead of using the integration constants $c_1$, $c_2$, $c_3$, $c_4$, $\mathcal{Q}_1$ and $\mathcal{Q}_2$, let us use $\tilde{c}_1$,  $\tilde{c}_2$,  $\tilde{c}_3$, $\tilde{c}_4$, $e_1$ and $e_2$ defined by the following relations:
\begin{align*}
\tilde{c}_1&=   \frac{b_1^{3/2} \, (a_2 b_1 - a_1 b_2)^{3/2}}{h_2^3(a_1 b_3 - a_3 b_1)} \,c_1    ,\\
\tilde{c}_2&=\frac{b_1^2(a_1 b_2-a_2 b_1)}{a_1 b_3-a_3 b_1}c_2-\frac{b_2^2\Lambda}{3(a_1 b_2-a_2 b_1)^2}\;,\\
\tilde{c}_3&=-\frac{b_1(a_1 b_2-a_2 b_1)^2}{h_2^2(a_1 b_3-a_3 b_1)}c_3-\frac{2b_2\Lambda}{3h_2^2(a_1 b_2-a_2 b_1)}\;,\\
\tilde{c}_4&=\frac{b_1^{3/2} \, (a_1 b_2 - a_2 b_1)^{3/2}}{h_2^3(a_1 b_3 - a_3 b_1)} \,c_4    ,  \\
e_1&=\frac{h_2^2 b_1(a_1 b_2-a_2 b_1)}{\sqrt{2\lambda}\,(a_1 b_3-a_3 b_1)} \mathcal{Q}_1\,,\\
e_2&=\frac{i h_2^2 b_1(a_1 b_2-a_2 b_1)}{\sqrt{2\lambda}\,(a_1 b_3-a_3 b_1)}  \mathcal{Q}_2\,.
\end{align*}
Along with this, we shall perform a coordinate transformation $(\tau, \sigma, x, y)\rightarrow (t, \phi,p,q)$ defined by
\begin{align*}
x^2&=b_1^{-1}\left(h_2^2\,p^2 +\frac{b_2}{a_1 b_2-a_2 b_1}\right)^{-1}-b_1^{-1}a_1\,, \nonumber \\
y^2&=b_1^{-1}\left( h_2^2\, q^2  -\frac{b_2}{a_1 b_2-a_2 b_1}\right)^{-1}+b_1^{-1}a_1\,, \\
\tau&=\frac{\sqrt{b_1^3(a_2 b_1-a_1 b_2)}}{h_1(a_1 b_3-a_3 b_1)}\left(\frac{a_1}{h_2 b_1}t + \frac{a_2}{h_2^3 (a_1 b_2-a_2 b_1)}\phi  \right)\,,\\
\sigma&=\frac{\sqrt{b_1^3(a_2 b_1-a_1 b_2)}}{h_1(a_1 b_3-a_3 b_1)}\left(\frac{1}{h_2}\,t  +\frac{b_2}{h_2^{3}(a_1 b_2-a_2 b_1)}\phi  \right)\,.
\end{align*}
Note that the constant parameters $h_1$ and $h_2$ that appear in the above redefinitions are not integration constants of the fields found in Sec. \ref{SubSec.geral}, rather they are arbitrary parameters that can be eliminated by a coordinate transformation along with the redefinition of some integration constants and have been introduced for sake of future convenience. In particular, $h_1$ and $h_2$ have non physical meaning.
Putting all these transformations together and using the constraint (\ref{c5-geral}), we eventually arrive at the following line element
\begin{multline*}
ds^2=-\frac{Q(q)}{h_1^2\rho^2}(dt-p^2d\phi)^2+\frac{\rho^2}{Q(q)}dq^2\\
+\frac{P(p)}{\rho^2}(dt+q^2d\phi)^2+ \frac{\rho^2}{h_1^2P(p)}dp^2\,,
\end{multline*}
with $\rho^2=p^2+q^2$ and $Q(q)$ and $P(p)$ being the following quartic polynomials
\begin{equation}\label{PQ}
  \left.
     \begin{array}{ll}
       P(p)&= h_1^{-2}\left(\frac{\tilde{c}_2}{ h_2^4}+\tilde{c}_1 p+\tilde{c}_3 p^2-\frac{\Lambda}{3} p^4\right)\,,\\
\\
Q(q)&= \frac{\tilde{c}_2}{ h_2^4} +e_1^2+e_2^2+\tilde{c}_4 q-\tilde{c}_3 q^2-\frac{\Lambda}{3} q^4\,.
     \end{array}
   \right.
\end{equation}
Note that the integration constants $a$'s and $b$'s are absent in the latter form of the metric, proving that they have no physical meaning. We have been left with the six integration constants $\tilde{c}_1$, $\tilde{c}_2$, $\tilde{c}_3$, $\tilde{c}_4$, $e_1$ and $e_2$. However, not all of these six constants are of physical relevance. Indeed, inspecting Eq. (\ref{PQ}), we can easily note that the constant  $\tilde{c}_2$ can be eliminated by choosing a proper value for the arbitrary parameter $h_2$. Thus, actually, we have just five physical parameters.

Now, in order to show that the present solution is the Kerr-Newman-NUT-(A)dS spacetime,  let us redefine the three physical constants $\tilde{c}_1$, $\tilde{c}_3$ and $\tilde{c}_4$ in terms of the new constants $a$, $l$ and $m$ by means of the following relations:
\begin{align*}
\tilde{c}_1 &= 2l+\frac{2}{3}a^2l\Lambda-\frac{8}{3}l^3\Lambda\,, \qquad \tilde{c}_4 = -2m\,,\\
\tilde{c}_3 &= \Lambda\left(\frac{1}{3}a^2+2l^2\right)-1\,.
\end{align*}
Then, we can use the freedom in the choice of the parameters $h_1$ and $h_2$ to set
\begin{equation*}
h_1=a \quad \textrm{and} \quad  h_2 = \tilde{c}_2^{1/4} (a^2-l^2)^{-1/4}(1 - \Lambda l^2)^{-1/4}   \,.
\end{equation*}
A further coordinate transformation $(t, \phi,p,q)\rightarrow (\tilde{t}, \tilde{\phi},\theta,r)$ defined by
\begin{align*}
t &=(a+l)^2\tilde{\phi}-a \tilde{t}\,, \qquad \phi=\tilde{\phi}\,,\\
p &=l+a \cos\theta\,,  \qquad \quad \; \; \; q=r\,,
\end{align*}
brings the line element to the following final form:
\begin{align*}
ds^2&=\frac{\rho^2}{Q}dr^2-\frac{Q}{\rho^2}\left[d\tilde{t}-\left(a\sin^2\theta+4l\sin^2\frac{\theta}{2}\right)d\tilde{\phi}\right]^2\\
&+ \frac{P}{\rho^2}\left[ad\tilde{t}-(r^2+(a+l)^2)d\tilde{\phi}\right]^2+\frac{\rho^2}{P}\sin^2\theta d\theta^2\,,
\end{align*}
where
\begin{align*}
\rho^2&= r^2+(l+a\cos\theta)^2,\\
P&= \sin^2\theta \left(1+\frac{4}{3}\Lambda a l \cos\theta+\frac{1}{3}\Lambda a^2 \cos^2\theta\right),\\
Q&= a^2-l^2+ e_1^2+ e_2^2-2m r+r^2\\
& \quad \quad -\Lambda  \left[(a^2-l^2)l^2+\left(\frac{1}{3}a^2+2l^2\right)r^2+\frac{1}{3}r^4\right]\,.
\end{align*}
This metric is precisely the Kerr-Newman-NUT-(A)dS solution as written in Ref. \cite{GrifPodol}, with the constants $m$, $a$ and $l$ being interpreted respectively as the mass, angular momentum per mass and NUT parameter, while $e_1$ and $e_2$ are the electric and magnetic charges. For a comparison, see Eq. (17) of Ref. \cite{GrifPodol} and the choice of parameters adopted in Sec. 4.2 of this reference.
Finally, in terms of the new coordinates and parameters, the gauge field for this case becomes
\begin{align*}
\mathcal{A} = &\frac{\Big[e_1(l+a \cos\theta)+ e_2 r\Big]}{\rho^2}\, d\tilde{t}\\
&- \frac{e_1\Big[r^2+(l+a)^2\Big](l+a \cos\theta)}{a \rho^2}\, d\tilde{\phi}\\
& - \frac{e_2 r\Big[(l+a)^2-(l+a \cos\theta)^2\Big]}{a \rho^2}\,  d\tilde{\phi}\,.
\end{align*}
In order to obtain this form for the gauge field $\mathcal{A}$, we have assumed $\lambda=\frac{1}{2}$, which is the coupling constant usually adopted for the electromagnetic field in the action (\ref{Action}).

\subsection{Case $P'_1\neq0$ and $P'_2=0$}

Now let us consider the case $P'_1\neq0$ and $P'_2=0$, tackled in Sec. \ref{SubSec.Pleb}. Here we will show that the general solution treated in this case represents a twisting but non-accelerating solution with vanishing rotation parameter inside the Pleba\'{n}ski-Demia\'{n}ski class of metrics  which contains, as a special case, the charged Taub-NUT-(A)dS spacetime. In order to demonstrate these assertions, let us consider the metric given by (\ref{LineElement}) with the functions $P_1$, $P_2$, $S_1$, $S_2$, $N_1$, $N_2$, $\Delta_1$ and $\Delta_2$ as given in Sec. \ref{SubSec.Pleb} along with the constraint (\ref{c5-Pleb}). Then, performing the coordinate transformation
\begin{align*}
\tau&=\frac{p_2}{\sqrt{a_1}}t-\frac{2\sqrt{a_2}}{\sqrt{a_1}}(p_2-1)\phi\,, \qquad x=p\,,\\
\sigma&=\frac{1}{\sqrt{a_1}}t-\frac{2\sqrt{a_2}}{\sqrt{a_1}}\phi\,,\qquad \qquad \quad \, \; y=r\,,
\end{align*}
and the redefinition of the integration constants
\begin{align*}
l&=\sqrt{a_2}\,,\qquad e_1=\sqrt{\frac{2}{a_1a_2\lambda}}\mathcal{Q}_1\,,\\
m&=-\frac{c_4}{2}\,,\qquad e_2=\sqrt{\frac{2}{a_1a_2\lambda}}\mathcal{Q}_2\,,
\end{align*}
one can check that the line element takes the form
\begin{align*}
ds^2=-Q\Big[dt-2l(1&-p)d\phi\Big]^2+\frac{dr^2}{Q}\\
&+(r^2+l^2)\left(P d\phi^2+\frac{dp^2}{P}\right)\,,
\end{align*}
with $P(p)$ and $Q(r)$ being the polynomials
\begin{align*}
P&=c_1 p^2+c_2 p+c_3\,,\\
Q&=\frac{1}{r^2+l^2}\Bigg[c_1 l^2+l^4\Lambda+e_1^2+e_2^2\Bigg.\\
&\qquad \qquad \qquad \; \; \Bigg.-2m r-(c_1+2l^2 \Lambda)r^2-\frac{\Lambda}{3}r^4\Bigg]\,.
\end{align*}
This line element represents a twisting but non-accelerating solution with vanishing rotation parameter inside the Pleba\'{n}ski-Demia\'{n}ski class of metrics, see Eq. (30) of Ref. \cite{GrifPodol} for a comparison.
With these changes of coordinates and parameters, the gauge field is now written as
\begin{equation*}
\mathcal{A}=\frac{(l^2-r^2)e_1+2e_2 l r}{2l(l^2+r^2)} \Big[dt-2l(1-p)d\phi\Big],
\end{equation*}
where we have set the coupling constant to be $\lambda=\frac{1}{2}$.
In particular, adopting $c_1=-1$, $c_2=0$ and $c_3=1$ and using the coordinate transformation $p = \cos\theta$, the latter metric above becomes
the charged Taub-NUT spacetime, see Eq. (31) of Ref. \cite{GrifPodol}.

\subsection{Case $P'_1=0$ and $P'_2=0$}

In this section we are going to show that for the case $P'_1=0$ and $P'_2=0$ our metric is either the Reissner-Nordstr\"{o}m spacetime with cosmological constant, when we follow the path $S_1'=0$ and $S_2'\neq0$, or the charged Nariai spacetime, when $S_1'=0$ and $S_2'=0$.

\subsubsection{The subcase $S_1'=0$ and $S_2'\neq0$}

For the subcase  $S_1'=0$ and $S_2'\neq0$, let us start with the metric (\ref{Metric-s1s2})
with the functions $S_1$, $S_2$, $N_1$, $N_2$ and $\Delta_2$ as defined in Sec. \ref{SubSec.s1cte} along with the constraint (\ref{d2-s1cte}).
 Then, using the new coordinates $(\tilde{t},\tilde{\phi},\theta,r)$ defined by
\begin{align*}
\phi&=\frac{1}{c_1\sqrt{a_1}}\tilde{\phi}\,,\qquad x=\frac{1}{c_1}(\theta-c_2)\,,\\
t&=\frac{1}{c_1\sqrt{a_2}}\tilde{t}\,,\qquad \; y=c_1 r\,,
\end{align*}
and redefining the integration constants as
\begin{equation*}
m=-\frac{d_1}{2c_1^3}\,, \;\;\;\; e_1=\frac{\mathcal{Q}_1}{c_1\sqrt{2a_1\lambda}}\,, \;\;\;\; e_2=\frac{\mathcal{Q}_2}{c_1^2\sqrt{2a_2\lambda}}\,,\\
\end{equation*}
it follows that the line element becomes
\begin{equation*}
ds^2=-f d\tilde{t}{\,}^2+f^{-1}dr^2+r^2\Big(d\theta^2+\sin^2\theta d\tilde{\phi}^2\Big),
\end{equation*}
where $f$ is the following function
\begin{equation*}
f(r)=1-\frac{2m}{r}+\frac{e_1^2+e_2^2}{r^2}-\frac{\Lambda}{3}r^2\,.
\end{equation*}
In this form, we easily recognise the metric as the Reissner-Nordstr\"{o}m solution \cite{GrifPodol}.
In these coordinates the gauge field takes the following simple form
\begin{equation*}
\mathcal{A}=\frac{e_2}{r}d\tilde{t}+e_1\cos\theta d\tilde{\phi}\,.
\end{equation*}

\subsubsection{The subcase $S_1'=0$ and $S_2'=0$}

For the subcase  $S_1'=0$ and $S_2'=0$, we shall start with the metric (\ref{Metric-s1s2})
with the functions $S_1$, $S_2$, $N_1$, $N_2$ and $\Delta_2$ as defined in Sec. \ref{SubSec.s1s2cte}, along with the constraint (\ref{d2-s1s2cte}).
In order to put this metric in a more convenient form, consider the new coordinates $(\tilde{t},\tilde{\phi},\theta,\tilde{y})$ defined by
\begin{align*}
t&=\frac{1}{\sqrt{a_1 a_2(2 b_2\Lambda-c_1^2)}}\tilde{t}\,,\qquad x=\frac{\theta-c_2}{c_1}\,,\\
y&=\frac{1}{\sqrt{2 b_2\Lambda-c_1^2}}(\tilde{y}-c_3)\,, \quad \; \; \, \phi=\frac{1}{c_1\sqrt{a_1}}\tilde{\phi}\,.
\end{align*}
Then, redefining the charges $\mathcal{Q}_1$ and $\mathcal{Q}_2$ as follows
\begin{equation*}
e_1=\frac{\mathcal{Q}_1}{\sqrt{2 a_1 b_2\lambda}}\,,\qquad e_2=\frac{\mathcal{Q}_2}{\sqrt{2 a_1 a_2 b_2\lambda}}\,,
\end{equation*}
it turns out that the constraint (\ref{d2-s1s2cte}) is written as
\begin{align*}
\frac{b_2}{c_1^2}=\frac{1-e_1^2+e_2^2}{\Lambda(1+2e_2^2)}\,.
\end{align*}
Adopting these new coordinates and charge parameters, the line element is given by
\begin{align*}
ds^2&=\frac{(1-e_1^2+e_2^2)}{\Lambda(1-2e_1^2)}\left(-\sin^2\tilde{y}\, d\tilde{t}{\,}^2+d\tilde{y}^2\right)\\
&\qquad \qquad \; \; +\frac{(1-e_1^2+e_2^2)}{\Lambda(1+2e_2^2)}\left(d\theta^2+\sin^2\theta d\tilde{\phi}^2\right)\,.
\end{align*}
As stressed out earlier, this space is just a product of the two-dimensional (anti-)de Sitter space with a sphere, the so-called charged (anti-)Nariai spacetime \cite{BatNariai}. Assuming the coupling constant to be $\lambda=\frac{1}{2}$, the gauge field for the new coordinates becomes
\begin{align*}
\mathcal{A}= \sqrt{1-e_1^2+e_2^2} &\left[\frac{e_2}{\sqrt{\Lambda(1-2e_1^2)}}\cos\tilde{y}d\tilde{t}\right.\\
&+\left.\frac{e_1}{\sqrt{\Lambda(1+2e_2^2)}}\cos\theta d\tilde{\phi}^2\right]\,.
\end{align*}

\end{document}